Entire latex paper,  no figures
to appear in Phys. Rev. C
\documentstyle[12pt]{article}
\begin{document}

\vspace{3cm}

\begin{center}

{\Large \bf Density Fluctuations \\ [5mm]
in the Quark-Gluon Plasma} \\

\vspace{3cm}

Stanis\l aw Mr\' owczy\' nski\footnote{E-mail: MROW@FUW.EDU.PL} \\[3mm]
\it
So\l tan Institute for Nuclear Studies,\\
ul. Ho\. za 69, PL - 00-681 Warsaw, Poland \\
and Institute of Physics, Pedagogical University,\\
ul. Konopnickiej 15, PL - 25-406 Kielce, Poland\\
\vspace{2cm}
\rm
\begin{minipage}{13cm}
\baselineskip=12pt

{\small \qquad Using the kinetic theory we discuss how the particle and 
energy densities of the quark-gluon plasma fluctuate in a space-time cell.
The fluctuations in the equilibrium plasma and in that one from the early 
stage of ultrarelativistic heavy-ion collisions are estimated. Within
the physically interesting values of the parameters involved the 
fluctuations appear sizeable in both cases.}

\end{minipage}
\end{center}

\vspace{0.5cm}

{\it PACS:} 12.38.M, 52.25.Gj, 25.75.+r

{\it Keywords:} Quark-gluon plasma; Fluctuations; Relativistic 
heavy-ion collisions

\vspace{2cm}

\begin{center}
{\it 23-th May 1997 \\}
{\it Revised 26-th August 1997 \\}
{\it 2-nd revision 24-th November 1997 \\}
\end{center}

\baselineskip=14pt
\newpage

The quark-gluon plasma, which is expected to be produced in energetic 
heavy-ion collisions, is usually described in terms of the averaged 
quantities like baryon or energy density, temperature, etc. However, 
it is often important to know how these quantities treated as local 
ones fluctuate around their average values. For example, it has 
been argued in our papers \cite{Mro97} that the color current fluctuating 
around zero can initiate the plasma instabilities. Gyulassy, Rischke and 
Zhang \cite{Gyu97} have recently shown that the hydrodynamical evolution of 
the quark-gluon plasma can be significantly influenced by the energy 
density fluctuations at the initial state. 

In this note we derive, using the kinetic theory methods, simple analytic 
expressions describing how the particle and energy densities fluctuate 
in a space-time cell of the volume $\Delta x^3 \Delta t$. Then, we use 
the derived formulas to estimate the particle and energy density 
fluctuations in the equilibrium plasma and in the parton system from 
the early stage of ultrarelativistic heavy-ion collisions at RHIC or LHC. 
Since the approach, which is employed, is rather simplistic our results 
should be mainly treated as a guiding for a more quantitative study in
the future. 

Let us start with the fluctuations of particle density $\rho (x)$, where 
$x \equiv (t, {\bf x})$ is a four-position. The average density is expressed 
through the distribution function $f$ as 
\begin{equation}\label{av-den}
\langle \rho (x) \rangle =\langle \rho \rangle 
= \int {d^3p \over (2\pi )^3} \; f({\bf p}) \;,
\end{equation}
where ${\bf p}$ is the particle momentum. The system is assumed be 
{\it on average} homogeneous and stationary and consequently, the
distribution function is independent of $x$. The momentum distribution 
is arbitrary.

The density correlation function for the classical system of noninteracting 
particles is given by a well known formula, see e.g. \cite{Akh75}, as 
\begin{equation}\label{den-cor-x}
A(x) \buildrel \rm def \over = 
\langle \rho (x_1) \rho (x_2) \rangle - \langle \rho \rangle^2 
= \int {d^3p \over (2\pi )^3} \; f({\bf p}) \;
\delta^{(3)} ({\bf x} -{\bf v}_p t) \;,
\end{equation}
where $x\equiv x_1-x_2 =(t_1-t_2,{\bf x}_1-{\bf x}_2)$ and ${\bf v}_p$ 
is the particle velocity equal ${\bf p}/E_p$ with 
$E_p \equiv \sqrt{{\bf p}^2 + m^2}$. Due to the average space--time 
homogeneity the correlation function $A(x)$ depends on the difference 
of $x_1$ and $x_2$ only. The space-time points $(t_1,{\bf x}_1)$ 
and $(t_2,{\bf x}_2)$ are correlated in the system of noninteracting 
particles if the particles fly from $(t_1,{\bf x}_1)$ to $(t_2,{\bf x}_2)$. 
For this reason the delta  $\delta^{(3)} ({\bf x} - {\bf v} t)$ is 
present in the formula (\ref{den-cor-x}). The momentum integral of the
distribution function simply represents the summation over particles.
Applicability of our classical approach to a quantum system such as 
a quark-gluon plasma is considered below. We also discuss there when 
the partons can be taken as noninteracting as in eq.~(\ref{den-cor-x}).

We get the particle density fluctuations in a given space-time cell 
averaging the correlation function $A(x)$ over the cell volume. 
Specifically, 
\begin{eqnarray}\label{coarse-grain-1}
\langle \rho^2 \rangle - \langle \rho \rangle^2 \equiv
\int d^4x \, \Delta (x) \, A(x) \;,
\end{eqnarray}
where the coarse-graining function $\Delta (x)$ is chosen to be of the 
gaussian shape i.e.
$$
\Delta (x) = {1 \over 4 \pi^2 \, \Delta t \, \Delta x^3} \;
{\rm exp}\bigg( -{t^2 \over 2 \Delta t^2} 
            -{{\bf x}^2 \over 2 \Delta x^2} \bigg) \;,
$$
with $\Delta t$ and $\Delta x$ denoting the root mean square of the 
cell size in time and space direction, respectively. 

Substituting the correlation function (\ref{den-cor-x}) into eq. 
(\ref{coarse-grain-1}) we get
\begin{eqnarray}\label{coarse-grain-2}
\langle \rho^2 \rangle - \langle \rho \rangle^2 =  
{1 \over (2\pi)^{3/2} \, \Delta x^2} \int {d^3p \over (2\pi )^3} \, 
{f({\bf p}) \over \sqrt{ \Delta x^2 + {\bf v}_p^2 \Delta t^2}}\;.
\end{eqnarray}
As seen, the fluctuations remain finite for $\Delta t = 0$ but 
grow to infinity when $\Delta x \rightarrow 0$.

For massless particles ${\bf v}_p^2=1$ and the formula (\ref{coarse-grain-2}) 
essentially simplifies. The relative fluctuations then are
\begin{eqnarray}\label{rel-par-den-flu}
{\langle \rho^2 \rangle  - \langle \rho \rangle^2 \over 
\langle \rho \rangle^2} = {1 \over (2\pi)^{3/2} \; \Delta x^2 \,
\sqrt{ \Delta x^2 + \Delta t^2}} \; {1 \over \langle \rho \rangle} \;,
\end{eqnarray}
and depend only on the average density and the size of the space-time cell. 
In the case of massive particles, a similar simplification occurs for
$\Delta t=0$.

One easily generalizes the above considerations to the energy density
fluctuations. The average energy density and the respective correlation 
function read
\begin{equation}\label{av-en-den}
\langle \varepsilon \rangle 
= \int {d^3p \over (2\pi )^3} \;E_p \: f({\bf p}) \;,
\end{equation}
$$
W(x) \buildrel \rm def \over = 
\langle \varepsilon (x_1) \varepsilon (x_2) \rangle 
- \langle \varepsilon \rangle^2 
= \int {d^3p \over (2\pi )^3} \;E_p^2\: f({\bf p}) \;
\delta^{(3)} ({\bf x} -{\bf v}_p t) \;.
$$
One finds the energy density fluctuations in the space-time cell as
\begin{eqnarray*}
\langle \varepsilon^2 \rangle - \langle \varepsilon \rangle^2 \equiv
\int d^4x \, \Delta (x) \, W(x) =
{1 \over (2\pi)^{3/2} \, \Delta x^2} \int {d^3p \over (2\pi )^3} \, 
{E_p^2 \: f({\bf p}) \over \sqrt{ \Delta x^2 + {\bf v}_p^2 \Delta t^2}}\;.
\end{eqnarray*}
For massless particles the relative energy density fluctuations are
\begin{eqnarray}\label{rel-en-den-flu}
{\langle \varepsilon^2 \rangle - \langle \varepsilon \rangle^2 
\over \langle \varepsilon \rangle^2 } =  
{1 \over (2\pi)^{3/2} \Delta x^2 \,\sqrt{ \Delta x^2 + \Delta t^2}} \;
{\langle e^2 \rangle \over \langle \varepsilon \rangle^2} \;,
\end{eqnarray}
with
\begin{equation}\label{av-en2-den}
\langle e^2 \rangle \equiv  \int {d^3p \over (2\pi )^3} \, 
E_p^2 \: f({\bf p}) \;.
\end{equation}
One sees that in contrast to the relative particle density fluctuations
given by eq. (\ref{rel-par-den-flu}), the energy density fluctuations 
(\ref{rel-en-den-flu}) depend not only on the average energy density 
but on the on the energy second moment, i.e. $\langle e^2 \rangle$, as well.

As already mentioned, our approach is classical. There are two quantum
effects: particle localization and bosonic or fermionic statistics which 
should be considered before the approach is applied to a quantum system. 
If the cell size $\Delta x$ is much larger than the average length of 
particle de Broglie wave $\lambda$, i.e. 
\begin{equation}\label{class}
\Delta x \gg \lambda \;,
\end{equation}
the particles can be treated as well localized. One also finds analogous 
requirement for $\Delta t$, but for the ultrarelativistic system under
consideration, the space and time scales, as the momentum and energy ones,
are usually of the same order. Thus, we further discuss only the condition 
(\ref{class}), which is checked below for the two cases: the equilibrium 
quark-gluon plasma and the parton system from the early stage of 
ultrarelativistic heavy-ion collisions. 

One takes into account the effect of quantum statistics substituting 
the distribution function $f$ by $f (1 \pm f)$, where $+$ is for bosons 
and $-$ for fermions, in eq.~(\ref{den-cor-x}). For the equilibrium 
quark-plasma the numerical 
effect of the statistics is small because the quark (fermionic) and gluon 
(bosonic) corrections, which are of comparable value, act in the opposite 
directions. In the case of the parton system from the early stage of 
ultrarelativistic heavy-ion collisions, where gluons dominate, we also 
neglect the effect of bosonic statistics. The correction is expected
to be small but significantly complicates the computation. The point 
is that the parton density is large at the early stage of the collision 
but the phase-space density, which matters for the quantum statistics
effects, is reduced due to the large longitudinal momentum range of 
the partons.

The quarks and gluons are taken as noninteracting in our approach. 
More specifically, the partons are assumed to follow the straight 
line trajectories in eq.~(\ref{den-cor-x}). Since the QCD color forces 
are of long range, the quarks and gluons always interact in a many-parton 
system. However, the effect of interaction is minor at the space 
scale which is much smaller than the inverse momentum transfer due 
to the interaction. Therefore, our formulas (\ref{rel-par-den-flu}) 
and (\ref{rel-en-den-flu}) are basically correct as long as 
\begin{equation}\label{free}
\Delta x \ll {1 \over q} \;,
\end{equation}
where $q$ is the characteristic momentum transfers discussed below. 

To obtain our final formulas (\ref{rel-par-den-flu}) and 
(\ref{rel-en-den-flu}) the partons have been assumed to be massless.
This assumption is correct for the case of equilibrium plasma at the
sufficiently large temperature when the interaction can be treated 
as a small perturbation. The situation is less clear when the early 
stage of nucleus-nucleus collision is considered. Then, the partons 
are off mass-shell. Nevertheless we still treat them as massless i.e.
we assume that the parton mass (`offeshellness') is smaller than 
$1/\Delta x$. We adopt this assumption keeping in mind that the partons 
with the `offshellness' $\mu$ decay after $\mu^{-1}$. Thus, they can be 
treated as noninteracting at a much smaller scale only.

After these comments let us discuss the equilibrium quark-gluon plasma. 
The distribution function then reads
\begin{equation}\label{equi}
f^{\rm eq}({\bf p}) = { g_g \over e^{\beta E_p} - 1} +
{ g_q \over e^{\beta E_p} + 1} \;,
\end{equation}
where $\beta^{-1} = T$ is the temperature and $g_g=16$ and $g_q=24$ are 
the numbers of the internal degrees of freedom of quarks (of two flavours) 
and gluons within the SU(3) gauge group. The plasma is assumed to be 
baryonless and the quarks, as gluons, are massless. 

Substituting the function (\ref{equi}) to eqs. (\ref{av-den},\ref{av-en-den})
and (\ref{av-en2-den}) we get, respectively,
\begin{eqnarray*}
\langle \rho \rangle = & {34 \zeta(3) \over \pi^2} \: T^3& 
\cong 4.14 \: T^3 \;,\\[2mm]
\langle \varepsilon \rangle = &{37 \pi^2 \over 30} \: T^4& 
\cong 12.2 \: T^4 \;,\\[2mm]
\langle e^2 \rangle = &{462 \zeta(5) \over \pi^2} \: T^5& 
\cong 48.5 \: T^5 \;,
\end{eqnarray*}
with $\zeta (z)$ being the zeta Riemann function; $\zeta (3) \cong 1.202$
and $\zeta (5) \cong 1.037$. The relative fluctuations of particle and 
energy densities (\ref{rel-par-den-flu}) and (\ref{rel-en-den-flu}) then are 
\begin{eqnarray}
\sqrt{{\langle \rho^2 \rangle  - \langle \rho \rangle^2 \over 
\langle \rho \rangle^2}} &\cong& {0.124\over \Delta x^{3/2}\:T^{3/2}} \;, 
\nonumber\\
\sqrt{{\langle \varepsilon^2 \rangle - \langle \varepsilon \rangle^2 
\over \langle \varepsilon \rangle^2}} &\cong&  
{0.144\over \Delta x^{3/2}\:T^{3/2}} \;,\label{en-den-flu-eq} 
\end{eqnarray}
where we put for simplicity $\Delta t = 0$.

We estimate the average length of particle de Broglie wave as the inverse 
average momentum, which in turn is defined as 
$\sqrt{\langle\langle {\bf p}^2 \rangle\rangle}$, where 
$\langle\langle ... \rangle\rangle$ denotes averaging over particles. 
Therefore,
$$
\lambda = {1 \over \sqrt{\langle\langle {\bf p}^2\rangle\rangle}} 
= \sqrt{\langle \rho \rangle \over \langle e^2 \rangle} 
\cong {0.292 \over T} \;,
$$
and the condition (\ref{class}) gets the form $\Delta x \: T \gg 0.3$.

In the equilibrium quark-gluon plasma the characteristic momentum transfer 
can be identified with the Debye screening mass $m_D$, which for two 
flavours and three colors equals $gT$ with $g$ being the QCD coupling 
constant, see e.g. \cite{Mro90}. Then, the condition (\ref{free}) is 
$\Delta x \:T \ll 1/g$. 

One sees that the two conditions, which allows us to treat partons as 
free and classical, can be fulfilled simultaneously if $g^{-1} \gg 1$
i.e. when the plasma is weakly interacting. Unfortunately, at the 
temperatures of order of a few hundreds MeV $g$ is not much smaller 
than unity. Therefore, $\Delta x$ should be close to $T^{-1}$ that the 
two conditions are not badly violated. 

The result (\ref{en-den-flu-eq}) can be helpful in choosing the physically 
reasonable size of the elementary fluid cell in the numerical hydrodynamical 
calculations. On one hand, the cell should small enough to get details of 
the density profiles, but on the other hand, the cell size should be 
sufficiently large to reduce the fluctuations. One sees from 
eq. (\ref{en-den-flu-eq}) that at $T = 200$ MeV and $\Delta x= 1$ fm the 
energy density deviates from the average value by about 14\%. The fluctuations
increase to 41\% for $\Delta x= 0.5$ fm. However, such a small cell
is at the border line of applicability of our classical approach. 

The size of the space-time cell is sometimes dictated by the characteristic
scale of the phenomenon under considerations. When the deconfinement phase 
transition is discussed, the cell size $\Delta x$ should be identified with 
the inverse confinement scale parameter $\Lambda_{QCD} \cong 200$ MeV. At 
the critical temperature $T_c = 150$ MeV the energy density fluctuations 
are about 22\%. Such sizeable fluctuations can significantly speedup the
process of hadronization. However, one should keep in mind that near the 
phase transition the quark-gluon plasma is no longer a system of weakly 
interacting particles and the formulas derived above can provide only a very 
rough estimate.

Let us now consider the nonequilibrium plasma from the early stage of
ultrarelativistic heavy-ion collisions. It is commonly believed that 
the perturbative processes, which are under theoretical control, play 
an important, if not dominant, role in these collisions, see e.g. \cite{Gei95}. 
We adopt this conventional point of view and discuss density fluctuations
of hard and semihard partons which are characterized by the transverse 
momentum being relatively large, at least 2 GeV, when compared to the
QCD scale parameter $\Lambda_{QCD}$. 

We estimate the particle (mostly gluons) density taking the numbers from 
\cite{Bir92}, where it has been found that about 570 perturbative gluons 
are generated at the early stage of the central Au--Au collision at RHIC 
($\sqrt s = 200$ GeV per N--N collision) and 8100 at LHC ($\sqrt s = 6$ TeV 
per N--N collision). Assuming that all these gluons appear in the cylinder 
of the volume $\pi r_0^2 A^{2/3} l$, where $r_0 = 1.1$ fm, $A = 197$ and 
$l=1$ fm, we get the average densities
\begin{displaymath}
\langle \rho \rangle \cong \left\{ 
\begin{array}{ccl}
4.4 \;\; {\rm fm}^{-3} \;\;\;\;\; &{\rm for}&  \;\;\;\; {\rm RHIC} \;,\\  
63  \;\; {\rm fm}^{-3} \;\;\;\;\; &{\rm for}&  \;\;\;\; {\rm LHC} \;,
\end{array}   \right.
\end{displaymath}
which immediately provide (via eq. (\ref{rel-par-den-flu})) the density 
fluctuations. 

To estimate the energy density fluctuations one needs the parton momentum 
distribution. We take it in the form which corresponds to the flat rapidity 
distribution in the interval $(-Y,Y)$ i.e.
\begin{equation}\label{f-flat-y}
f({\bf p}) = {1 \over 2Y} 
\Theta(Y - y) \Theta(Y + y) \; h(p_{\bot}) \;
{1 \over p_{\bot} \, {\rm ch}y}\;, 
\end{equation}
where $y$ and $p_{\bot}$ denote the parton rapidity and transverse momentum.
We do not specify the transverse momentum distribution $h(p_{\bot})$ because 
it is sufficient for our considerations to demand that the distribution 
(\ref{f-flat-y}) is strongly elongated along the beam axis i.e. $e^Y \gg 1$.

The QCD-based computations, see e.g. \cite{Gei95}, show that the rapidity 
distribution of partons produced at the early stage of heavy-ion collisions 
is essentially gaussian with the width of about one to two units. When the 
distribution (\ref{f-flat-y}) is used to simulate the gaussian one, $Y$ does 
not measure the size of the `plateau' but rather the range over which the 
partons are spread. If one takes the gaussian distribution of the variance 
$\sigma$ and the distribution (\ref{f-flat-y}) of the same variance, then 
$Y = \sqrt{3} \, \sigma $. 

One computes the average energy (\ref{av-en-den}) and energy squared 
(\ref{av-en2-den}) densities with the distribution (\ref{f-flat-y}) as
\begin{eqnarray*}
\langle \varepsilon \rangle &=& 
{{\rm sh}Y \over Y} \:\langle p_{\bot} \rangle \langle \rho \rangle 
\cong {e^Y \over 2Y} \:\langle p_{\bot} \rangle \langle \rho \rangle \;,
\\[3mm]
\langle e^2 \rangle &=& 
{{\rm sh}2Y + 2Y\over 4Y} \: \langle p_{\bot}^2 \rangle \langle \rho \rangle
\cong {e^{2Y} \over 8Y} \: \langle p_{\bot} \rangle^2 \langle \rho \rangle \;,
\end{eqnarray*}
and gets the energy density fluctuations
\begin{eqnarray}\label{en-den-flu-neq}
\sqrt{{\langle \varepsilon^2 \rangle - \langle \varepsilon \rangle^2 
\over \langle \varepsilon \rangle^2}} \cong
{0.178\over \Delta x^{3/2}} \sqrt{{ Y \over\langle \rho \rangle}} \;,
\end{eqnarray}
where $\Delta t = 0$.

Taking the values of $Y$ given in \cite{Bir92} i.e. $Y \cong 2.5$ at RHIC 
and $Y \cong 5.0$ at LHC and the gluon density, which has been estimated 
above, one gets the energy density fluctuations. Eq. (\ref{en-den-flu-neq})
tells us that the fluctuations at RHIC are 13\% for $\Delta x = 1$ fm
and increase to 38\% for $\Delta x = 0.5$ fm. At LHC the fluctuations 
are smaller by factor 2.7. Since the average transverse momentum of 
perturbative partons is of GeV order and the longitudinal momentum is 
even larger, their wavelength is a small fraction of 1 fm. The condition 
(\ref{class}) is then easily satisfied for the cell size of interest. 
Neglecting of the interaction seems to be reasonable as well, when
the partons with small `offshellness' are taken into account.

Let us summarize our considerations. We have derived the formulas which 
describe how the particle and energy densities fluctuate in a space-time 
cell of the volume $\Delta x^3 \Delta t$. These formulas, which get a very 
simple form for massless particles, have been applied to estimate the density
fluctuations in the quark-gluon plasma. We have considered the equilibrium 
plasma and the anisotropic parton system from the early stage of 
ultrarelativistic heavy-ion collisions at RHIC or LHC. In both cases
the fluctuations can be large within the reasonable values of the
parameters of interest. Sizeable density fluctuations are of physical
interest when the temporal evolution of the plasma system is studied.
As shown in \cite{Gyu97}, the hydrodynamics is then noticeably modified. 
The density fluctuations seem to be even more important when the 
hadronization is analysed or one considers the processes, such as the 
$J/\psi$ dissociation in the plasma \cite{Kar96}, which are strongly 
density dependent.

\vspace{1cm}
I am very grateful to Mark. I. Gorenstein for critical reading of
the manuscript.

\end{document}